\documentclass[12pt]{article}
\usepackage{graphicx}

\def \b{{\cal B}}
\def \bea{\begin{eqnarray}}
\def \beq{\begin{equation}}
\def \ca{{\cal A}}
\def \eea{\end{eqnarray}}
\def \eeq{\end{equation}}

\def \ok{\overline{K}^0}
\def \s{\sqrt{2}}
\def \st{\sqrt{3}}
\def \sx{\sqrt{6}}
\def \ta{\tilde{A}}
\def \tc{\tilde{C}}
\def \te{\tilde{E}}
\def \ttl{\tilde{T}}

\begin{document}
\rightline{EFI 08-05}
\rightline{arXiv:0803.2385}
\rightline{March 2008}
\bigskip
\centerline{\bf FLAVOR SYMMETRY AND DECAYS OF CHARMED MESONS}
\smallskip
\centerline{\bf TO PAIRS OF LIGHT PSEUDOSCALARS}
\bigskip

\centerline{Bhubanjyoti Bhattacharya\footnote{bhujyo@uchicago.edu} and
Jonathan L. Rosner\footnote{rosner@hep.uchicago.edu}}
\centerline{\it Enrico Fermi Institute and Department of Physics}
\centerline{\it University of Chicago, 5640 S. Ellis Avenue, Chicago, IL 60637}

\begin{quote}

New data on the decays of the charmed particles $D^0$, $D^+$, and
$D_s$ to pairs of light pseudoscalar mesons $P$ allow the testing of flavor
symmetry and the extraction of key amplitudes.  Information on relative
strong phases is obtained.  One sees evidence for the expected interference
between Cabibbo-favored and doubly-Cabibbo-suppressed decays in the differing
patterns of $D^0 \to K_{S,L} \pi^0$ and $D^+ \to K_{S,L} \pi^+$ decays.
\end{quote}

\section{Introduction}

The application of SU(3) flavor symmetry to charmed particle decays can shed
light on such questions as the strong phases of amplitudes in these decays.
Such strong phases are non-negligible even in $B$ decays to pairs of
pseudoscalar mesons ($P$), and can be even more important in $D \to PP$
decays.  In the present paper we shall extract strong phases from charmed
particle decays using SU(3) flavor symmetry, primarily the U-spin symmetry
involving the interchange of $s$ and $d$ quarks.  A preliminary version of
this work was presented in Ref.\ \cite{Bhattacharya:2007jc}.

We recall the diagrammatic approach to flavor symmetry in Section 2.  We then
treat Cabibbo-favored decays in Section 3, turning to singly-Cabibbo-suppressed
decays in Section 4 and doubly-Cabibbo-suppressed decays in Section 5.  We
mention some other theoretical approaches in Section 6, and conclude in Section
7.

\section{Diagrammatic amplitude expansion}

\begin{figure}[t]
\mbox{\includegraphics[width=0.46\textwidth]{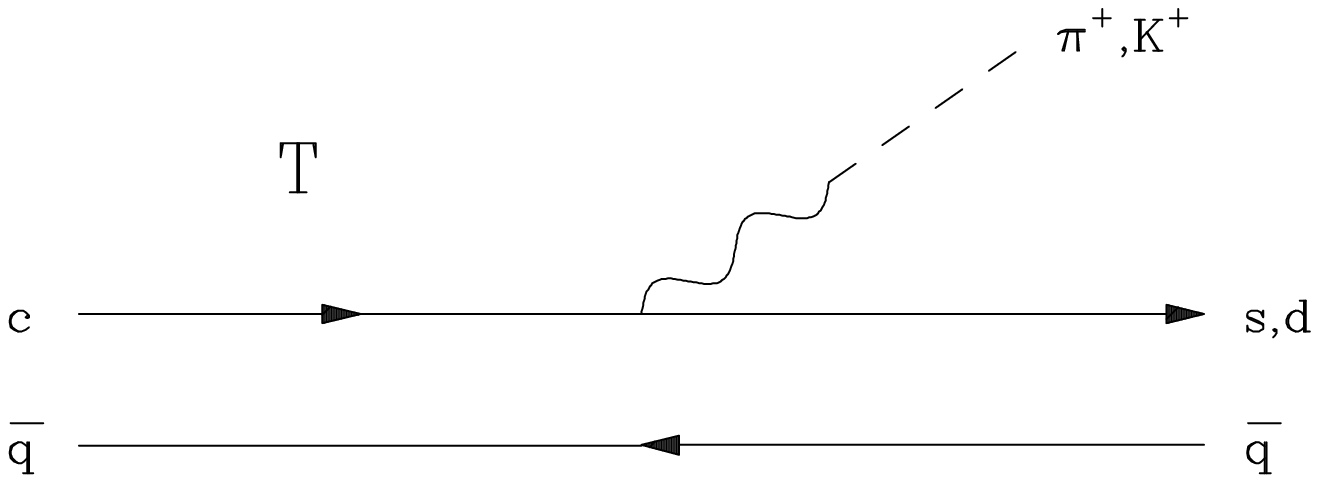} \hskip 0.3in
      \includegraphics[width=0.46\textwidth]{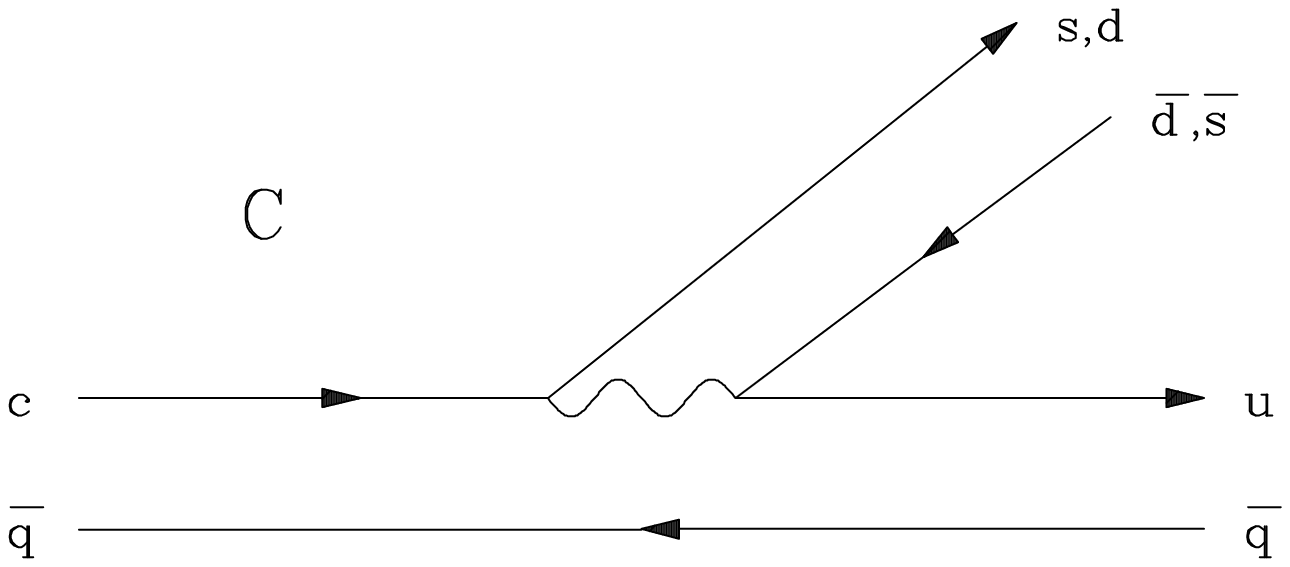}}
\vskip 0.3in
\mbox{\includegraphics[width=0.46\textwidth]{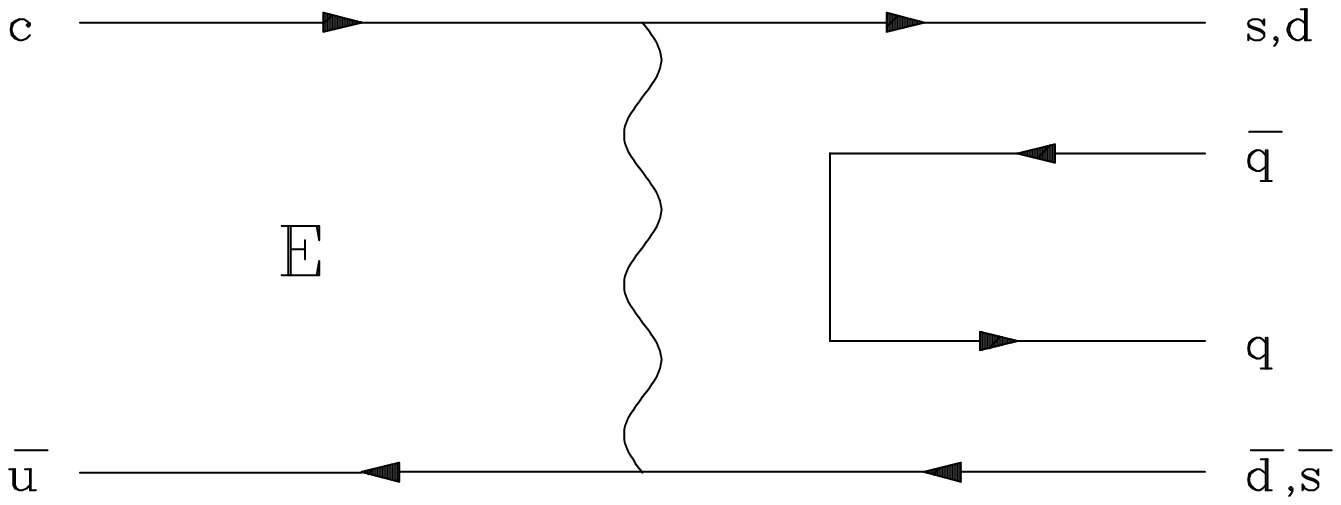} \hskip 0.3in
      \includegraphics[width=0.46\textwidth]{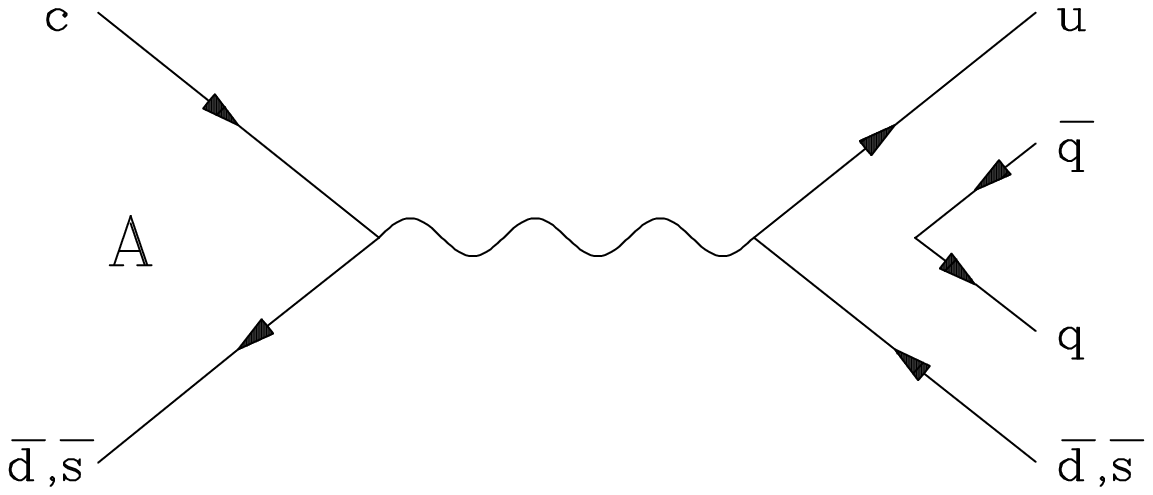}}

\caption{Flavor topologies for describing charm decays.  $T$: color-favored
tree; $C$: color-suppressed tree; $E$ exchange; $A$: annihilation.
\label{fig:TCEA}}
\end{figure}

We use a flavor-topology language for charmed particle decays
\cite{Chau:1983,Chau:1986}.  These topologies,
corresponding to linear combinations of SU(3)-invariant amplitudes, are
illustrated in Fig.\ \ref{fig:TCEA}.  Cabibbo-favored
(CF) amplitudes, proportional to the product $V_{ud} V^*_{cs}$ of
Cabibbo-Kobayashi-Maskawa (CKM) factors, will be denoted by unprimed
quantities; singly-Cabibbo-suppressed amplitudes proportional to $V_{us}
V^*_{cs}$ or $V_{ud} V^*_{cd}$ will be denoted by primed quantities; and
doubly-Cabibbo-suppressed quantities proportional to $V_{us} V^*_{cd}$ will
be denoted by amplitudes with a tilde.  The relative hierarchy of these
amplitudes is $1:\lambda:-\lambda:-\lambda^2$, where $\lambda = \tan \theta_C =
0.2317$ \cite{Yao:2006px,RS}.  Here $\theta_C$ is the Cabibbo angle.

% This is Figure 1

\section{Cabibbo-favored decays}

Amplitudes and their relative phases for Cabibbo-favored charm decays were
discussed in Ref.\ \cite{Rosner:1999}.  That analysis found large relative
phases of the $C$ and $E$ amplitudes relative to the dominant $T$ term, and an
approximate relation $A \simeq -E$.  An analysis \cite{Bhattacharya:2007jc}
based on the compilation in Ref.\ \cite{Yao:2006px} was consistent with this
conclusion.  The advent of new branching ratios for Cabibbo-favored $D_s$
decays \cite{Alexander:2008}, obtained independently of the branching ratio for
$D_s^+ \to \phi \pi^+$, changes this conclusion.  The relative phases of $C$
and $E$ with respect to $T$ are still large and their magnitudes are not
greatly changed, but now $A \simeq (-0.32 \pm 0.24)E$, in agreement with a
prediction $A \simeq -0.4 E$ in Ref.\ \cite{Gao:2007}.

In Table \ref{tab:CF} we show the results of extracting amplitudes ${\cal A} =
M_D[8 \pi {\cal B} \hbar/(p^* \tau)]^{1/2}$ from the branching ratios
${\cal B}$ \cite{Alexander:2008,Artuso:2008ri} and lifetimes $\tau$
\cite{Yao:2006px}.  Here $M_D$ is the mass of the decaying charmed particle,
and $p^*$ is the final c.m.\ 3-momentum.

% This is Table I
\begin{table}
\caption{Branching ratios \cite{Alexander:2008,Artuso:2008ri}, amplitudes, and
graphical representations for Cabibbo-favored charmed particle decays.
\label{tab:CF}}

\begin{center}
\begin{tabular}{|c|c|c|c|c|c|c|} \hline
Meson &    Decay    &     $\b$      & $p^*$ &    $|\ca|$    & Rep. &
Predicted \\
      &    mode     &     (\%)      & (MeV) &($10^{-6}$ GeV)&      &
$\b$ (\%) \\ \hline
$D^0$ & $K^- \pi^+$ & 3.891$\pm$0.077 & 861.1 & 2.52$\pm$0.02 & $T+E$ &
3.90 \\
      & $\ok \pi^0$ & 2.238$\pm$0.109 & 860.4 & 1.91$\pm$0.05 & $(C-E)/\s$ &
2.21 \\
      & $\ok \eta$  & 0.76$\pm$0.11 & 771.9 & 1.18$\pm$0.09 & $C/\st$ &
0.76 \\
      & $\ok \eta'$ & 1.87$\pm$0.28 & 564.9 & 2.16$\pm$0.16 & $-(C+3E)/\sx$ &
1.95 \\
\hline
$D^+$ & $\ok \pi^+$ & 2.986$\pm$0.067 & 862.4 & 1.39$\pm$0.02 & $C+T$ &
2.99 \\ \hline
$D_s^+$& $\ok K^+$  & 2.98$\pm$0.17 & 850.3 & 2.12$\pm$0.06 & $C+A$ &
3.02 \\
      & $\pi^+\eta$ & 1.58$\pm$0.21 & 902.3 & 1.50$\pm$0.10 & $(T-2A)/\st$ &
1.47 \\
      & $\pi^+\eta'$& 3.77$\pm$0.39 & 743.2 & 2.55$\pm$0.13 &$2(T+A)/\sx$ &
3.61 \\
\hline
\end{tabular}
\end{center}
\end{table}

% This is Figure 2
\begin{figure}
\begin{center}
\includegraphics[width=0.80\textwidth]{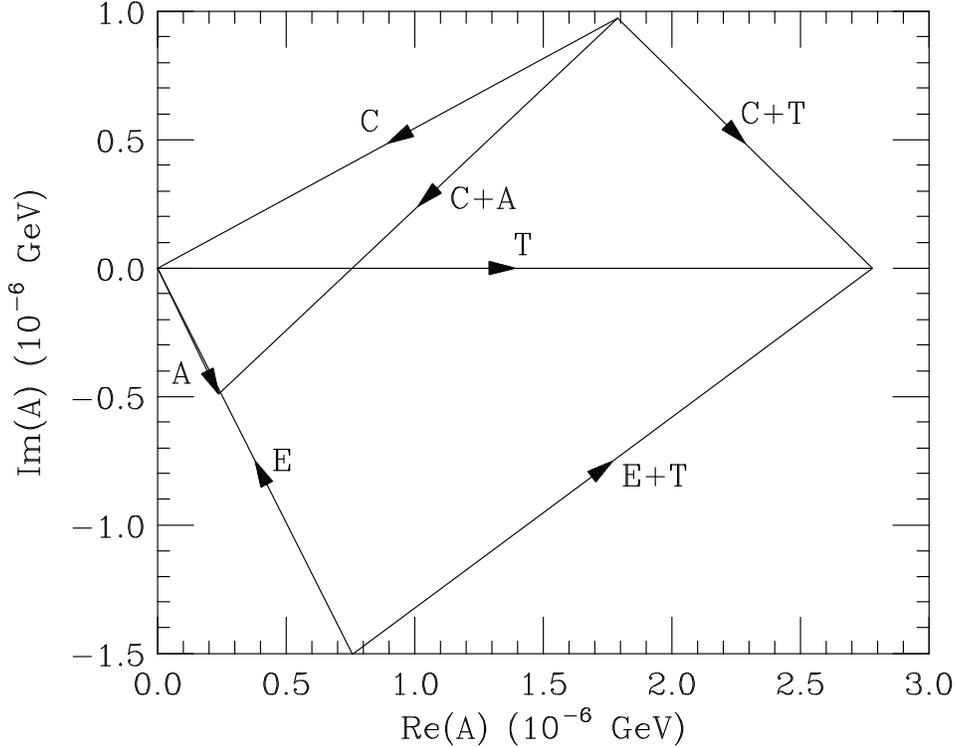}
\end{center}
\caption{Construction of Cabibbo-favored amplitudes from observed processes.
The sides $C+T$, $C+A$, and $E+T$ correspond to measured processes;
the magnitudes of other amplitudes listed in Table \ref{tab:CF} are also
needed to specify $T$, $C$, $E$, and $A$.
\label{fig:cf}}
\end{figure}

The extracted amplitudes, with $T$ defined to be real, are, in units of
$10^{-6}$ GeV:
\bea
T & = &  2.78 \pm 0.13 \\
C & = & (2.04 \pm 0.17)\exp[i(-151.5 \pm 1.7)^\circ]~;\\
E & = & (1.68 \pm 0.12)\exp[i(116.7 \pm 3.6)^\circ]~; \\
A & = & (0.55 \pm 0.39)\exp[i(-64^{+32}_{-8})^\circ]~.
\eea
These values update those quoted in Refs.\ \cite{Bhattacharya:2007jc} and
\cite{Rosner:1999}.  The amplitudes are shown on an Argand diagram in Fig.\
\ref{fig:cf}.  The fit has $\chi^2 = 0.64$ for one degree of freedom.  These
results are also obtained in Ref.\ \cite{Artuso:2008ri}.  Slightly different
amplitudes are obtained if one uses all measured branching ratios except
that for $D_s^+ \to \ok K^+$ as inputs, as in Ref.\ \cite{Chiang:2003}.
This method is algebraically convenient as one can eliminate an interference
term between $T$ and $A$ with a suitable combination of $D_s^+ \to \pi^+ \eta$
and $D_s^+ \to \pi^+ \eta'$ decay rates.  The predicted branching ratio,
$\b(D_s^+ \to \ok K^+) = 3.39 \%$, is in satisfactory agreement with experiment.

\section{Singly-Cabibbo-suppressed decays}

\subsection{SCS decays involving pions and kaons}

We show in Table \ref{tab:scskpi} the branching ratios, amplitudes, and
representations in terms of reduced amplitudes for singly-Cabibbo-suppressed
(SCS) charm decays involving pions and kaons.  The ratio of primed (SCS)
to unprimed (CF) amplitudes is assumed to be $\tan \theta_C = 0.2317$.
One then finds, in units of $10^{-7}$ GeV,
\bea
T' & = & 6.44~; \\
C' & = & -4.15 - 2.25i~; \\
E' & = & -1.76 + 3.48i~; \\
A' & = &  0.55 - 1.14i~.
\eea

The deviations from flavor SU(3) in Table \ref{tab:scskpi} are well known.  One
predicts $\b(D^0 \to \pi^+ \pi^-)$ larger than observed and $\b(D^0 \to K^+
K^-)$ smaller than observed.  One can account for some of this discrepancy via
the ratios of decay constants $f_K/f_\pi = 1.2$ and form factors $f_+(D \to
K)/f_+(D \to \pi) > 1$.  Furthermore, one predicts $\b(D^0 \to \pi^0 \pi^0)$
larger than observed and $\b(D^+ \to \pi^+ \pi^0)$ smaller than observed,
which means that the $\pi \pi$ isospin triangle [associated with the fact that
there are two independent amplitudes with $I=(0,2)$ for three decays] has a
different shape from that predicted by rescaling the CF amplitudes.  One
predicts equal decay amplitudes for $D^+ \to K^+ \ok$ and $D_s \to \pi^+ K^0$;
the experimental branching ratio for the former is about 20\% above the
predicted value.

% This is Table II
\begin{table}
\caption{Branching ratios, amplitudes, decomposition in terms of reduced
amplitudes, and predicted branching ratios for singly-Cabibbo-suppressed (SCS)
charm decays involving pions and kaons.
\label{tab:scskpi}}
\begin{center}
\begin{tabular}{|c|c|c|c|c|c|c|} \hline
Meson &    Decay    &         $\b$         & $p^*$ &    $|\ca|$    &
   Rep.    &   Predicted    \\
      &    mode     &     $(10^{-3})$      & (MeV) &($10^{-7}$ GeV)&
           & $\b~(10^{-3})$ \\ \hline
$D^0$ &$\pi^+ \pi^-$& 1.37$\pm$0.03$^{~a}$ & 921.9 & 4.57$\pm$0.05 &
$-(T'+E')$ &    2.23 \\
      &$\pi^0 \pi^0$& 0.79$\pm$0.08$^{~a}$ & 922.6 & 3.46$\pm$0.18 &
$-(C'-E')/\s$ & 1.27 \\
      &  $K^+ K^-$  & 3.93$\pm$0.07$^{~b}$ & 791.0 & 8.35$\pm$0.08 &
$(T'+E')$  &    1.92 \\
      &  $K^0 \ok$  & 0.37$\pm$0.06$^{~b}$ & 788.5 & 2.57$\pm$0.35 &
    0      &   0     \\ \hline
$D^+$ &$\pi^+ \pi^0$& 1.28$\pm$0.08$^{~a}$ & 924.7 & 2.77$\pm$0.09 &
$-(T'+C')/\s$ & 0.87 \\
      &  $K^+ \ok$  & 6.17$\pm$0.20$^{~b}$ & 792.6 & 6.58$\pm$0.11 &
$T'-A'$    & 5.12 \\ \hline
$D_s^+$&$\pi^+ K^0$ & 2.44$\pm$0.30$^{~c}$ & 915.7 & 5.84$\pm$0.36 &
$-(T'-A')$ & 2.56 \\
      & $\pi^0 K^+$ & 0.75$\pm$0.28$^{~c}$ & 917.1 & 3.24$\pm$0.60 &
$-(C'+A')/\s$ & 0.87 \\
\hline
\end{tabular}
\end{center}
\leftline{$^a$ From Ref.\ \cite{Yao:2006px}; $^b$ Ref.\ \cite{Bonvicini:2007}
averaged with Ref.\ \cite{Yao:2006px}; $^c$ Ref.\ \cite{Alexander:2008}
combined with \cite{Adams:2007mx}.}
\end{table}

The decay $D^0 \to K^0 \ok$ is forbidden by SU(3); the branching ratio of
$2 {\cal B}(D^0 \to K_S^0 K_S^0) = (2.92 \pm 0.64 \pm 0.18) \times
10^{-4}$ reported by CLEO \cite{Bonvicini:2007} is more than a factor of two
below the average in Ref.\ \cite{Yao:2006px}.  Estimates of SU(3)-breaking
effects lead to predictions for $\b(D^0 \to K^0 \ok)$ ranging from a few parts
in $10^4$ \cite{Eeg:2001un} to $3 \times 10^{-3}$ \cite{Pham:1987}.

\subsection{SCS decays involving $\eta,\eta'$}

The amplitudes $C$ and $E$ extracted from Cabibbo-favored charm decays imply
values of $C' = \lambda C$ and $E' = \lambda E$ which may be used in
constructing amplitudes for singly-Cabibbo-suppressed $D^0$ decays involving
$\eta$ and $\eta'$.  In Table \ref{tab:scseta} we write amplitudes multiplied
by factors so that they involve unit coefficient of an amplitude $SE'$
describing a disconnected ``singlet'' exchange amplitude for $D^0$ decays
\cite{Chiang:2003}.  Similarly the decays $D^+ \to (\pi^+ \eta, \pi^+ \eta')$
and $D_s^+ \to (K^+ \eta, K^+ \eta')$ may be described in terms of a
disconnected singlet annihilation amplitude $SA'$, written with unit
coefficient in Table \ref{tab:scseta}.  For experimental values we have used
new CLEO measurements as reported in Ref.\ \cite{Adams:2007mx}.
(See Table \ref{tab:pluseta}.)

% This is Table III
\begin{table}
\caption{Real and imaginary parts of amplitudes for SCS charm decays involving
$\eta$ and $\eta'$, in units of $10^{-7}$ GeV as
predicted in Ref.\ \cite{Chiang:2003}.
\label{tab:scseta}}
\begin{center}
\begin{tabular}{|c|c|r|r|c|} \hline
Amplitude   & Expression & Re & Im & $|\ca_{\rm exp}|$ \\ \hline
$-\sx \ca(D^0 \to \pi^0 \eta)$ & $2E'-C' +SE'$
     & 0.63 & 9.21 & $7.79 \pm 0.54$ \\
$\frac{\st}{2} \ca(D^0 \to \pi^0 \eta\,')$ & $\frac12(C' + E') + SE'$
     & $-2.95$ & 0.62 & $3.54 \pm 0.35$ \\
$\frac{3}{2 \s} \ca(D^0 \to \eta \eta)$ & $C'+ SE'$
     & $-4.14$ & $-2.25$ & $5.91 \pm 0.34$ \\
$-\frac{3 \s}{7} \ca(D^0 \to \eta \eta\,')$ & $\frac17(C' + 6E')+SE'$
     & $-2.10$ & 2.66 & $3.48 \pm 0.38$ \\
$\st \ca(D^+ \to \pi^+ \eta)$ & $T'+2C'+2A'+ SA'$
     & $-0.75$ & $-6.77$ & 8.21$\pm$0.26 \\
$-\frac{\sx}{4} \ca(D^+ \to \pi^+ \eta\,')$ & $\frac14(T'-C'+2A')+SA'$
     & 2.92 & $-0.01$ & 3.72$\pm$0.15 \\
$-\st \ca(D_s^+ \to \eta K^+)$ & $-(T'+2C')+SA'$
     & 1.85 & 4.50 & 8.05$\pm$0.88 \\
$\frac{\sx}{4} \ca(D_s^+ \to \eta\,' K^+)$ & $\frac14(2T'+C'+3A')+SA'$
     & 2.59 & $-1.41$ & 3.43$\pm$0.57 \\ \hline
\end{tabular}
\end{center}
\end{table}

% This is Figure 3
\begin{figure}
\mbox{\includegraphics[width=0.46\textwidth]{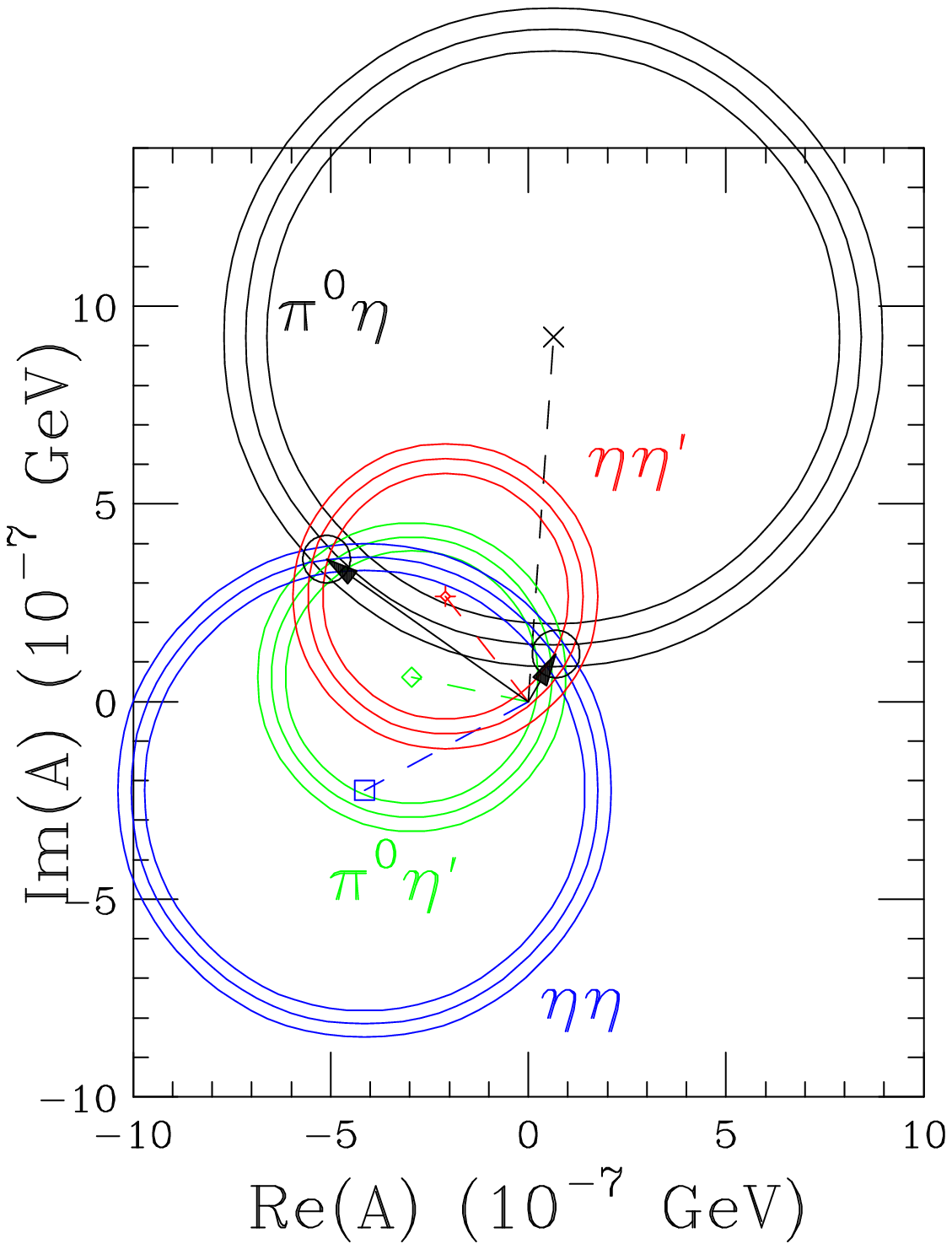}
      \includegraphics[width=0.46\textwidth]{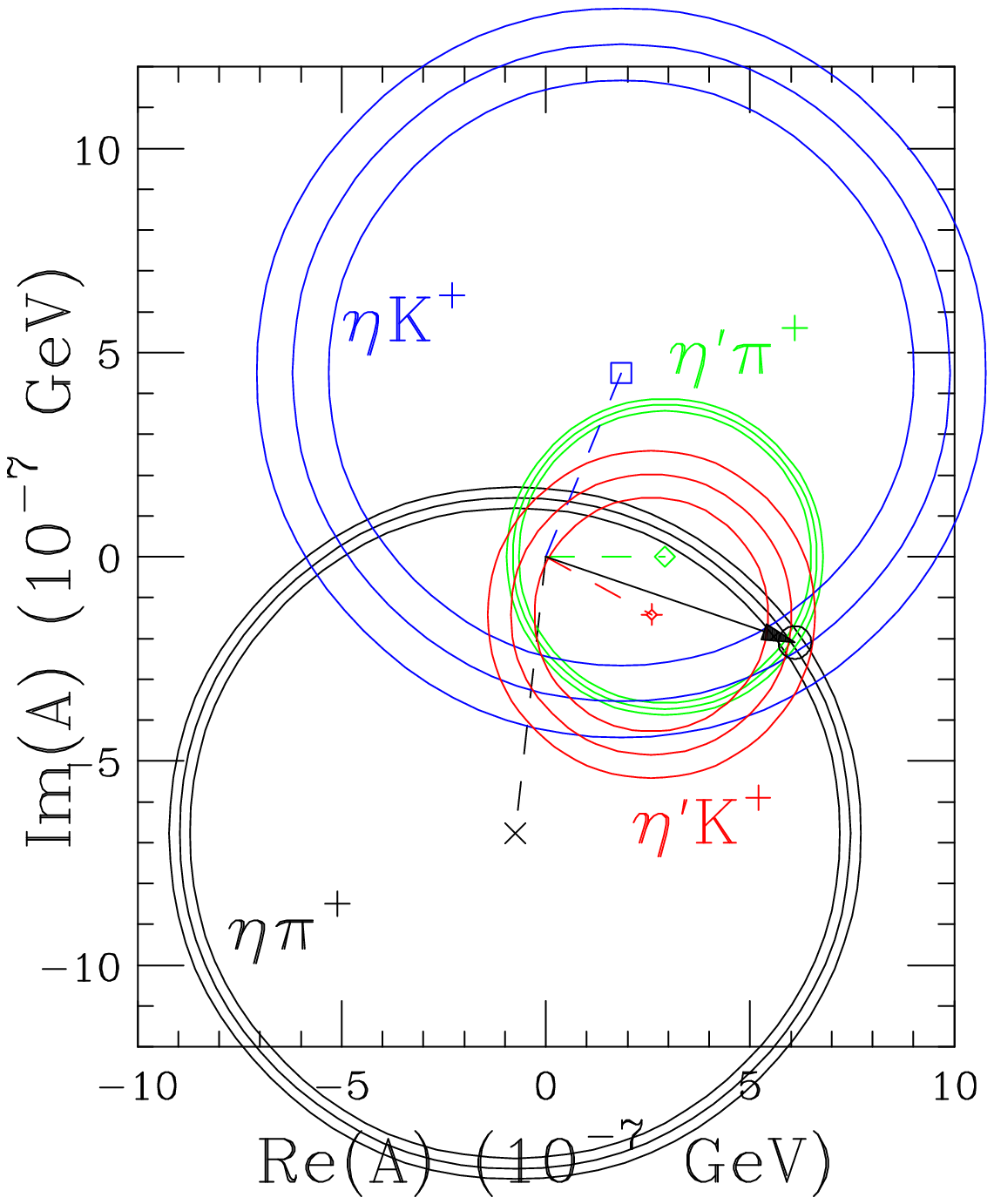}}
\caption{Graphical construction to obtain the disconnected singlet annihilation
amplitudes $SE'$ (left) and $SA'$ (right) from magnitudes of SCS $D^0$,
$D^+$, and $D_s^+$ decays involving $\eta$ and $\eta'$.  Left: Black: $D^0 \to
\pi^0 \eta$; green: $D^0 \to \pi^0 \eta'$; blue: $D^0 \to \eta \eta$; red:
$D^0 \to \eta \eta'$.  Right: Black:  $D^+ \to \eta \pi^+$.  Green: $D^+ \to
\eta' \pi^+$.  Blue: $D_s^+ \to \eta K^+$.  Red: $D_s^+ \to \eta' K^+$.  The
small circles with arrows pointing to them show the solution regions.  The
arrows denote the complex amplitudes $-SE'$ (left) and $-SA'$ (right).
\label{fig:sesa}}
\end{figure}

% This is Table IV
\begin{table}
\caption{Branching ratios and amplitudes for $D^0$, $D^+$, and $D_s^+$ SCS
decays involving $\eta$ and $\eta'$.
\label{tab:pluseta}}
\begin{center}
\begin{tabular}{|c|c|c|c|c|} \hline
Meson   &    Decay     &     $\b$      & $p^*$ &     $|\ca|$       \\
        &     mode     &  $(10^{-4})$  & (MeV) & ($10^{-7}$ GeV) \\ \hline
$D^0$   & $\pi^0 \eta$ & $6.10 \pm 0.85^{~a}$ & 846.2 & $3.18 \pm 0.22$ \\
        &$\pi^0 \eta'$ &  $8.1 \pm 1.6^{~b}$  & 678.0 & $4.09 \pm 0.41$ \\
        & $\eta \eta$  & $16.7 \pm 1.9^{~b}$  & 754.6 & $5.57 \pm 0.32$ \\
        & $\eta \eta'$ & $12.6 \pm 2.7^{~b}$  & 536.8 & $5.74 \pm 0.62$ \\
\hline
$D^+$   & $\pi^+ \eta$ & $34.3 \pm 2.1^{~a}$  & 848.4 & $4.74 \pm 0.15$ \\
        &$\pi^+ \eta'$ & $45.2 \pm 3.6^{~a}$  & 680.5 & $6.08 \pm 0.24$ \\
\hline
$D_s^+$ &  $K^+ \eta$  & $14.1 \pm 3.1^{~c}$  & 835.0 & $4.65 \pm 0.51$ \\
        &  $K^+ \eta'$ & $15.8 \pm 5.3^{~c}$  & 646.1 & $5.60 \pm 0.94$ \\
\hline
\end{tabular}
\end{center}
\leftline{$^a$ Average of Refs.\ \cite{Yao:2006px} and \cite{Artuso:2008ri}.
$^b$ Ref.\ \cite{Artuso:2008ri}. $^c$ Ref.\ \cite{Alexander:2008}
combined with \cite{Adams:2007mx}.}
\end{table}

We show in Fig.\ \ref{fig:sesa} the construction proposed in Refs.\
\cite{Chiang:2003} to obtain the amplitudes $SE'$ and $SA'$.  For $SE'$,
two solutions are found \cite{Artuso:2008ri}: in units of $10^{-7}$ GeV, $SE'
% \simeq -5.1 + 3.6i$
= (5.3 \pm 0.5) - i (3.5 \pm 0.5)$
and $SE'
% \simeq 0.7 + 1.2i$
= (-0.7 \pm 0.4) - i(1.0 \pm 0.6)$.
In the first, $|SE'|$ is uncomfortably large in
comparison with the ``connected'' amplitudes.  The only solution for $SA'
\simeq - 6.1 + 2.1i$ does not exhibit any suppression in comparison with the
connected SCS amplitudes.

\subsection{Sum rules for $D^0 \to (\pi^0 \pi^0, \pi^0 \eta, \eta \eta, \pi^0
\eta', \eta \eta')$}
It appears from the representations of the Cabibbo-suppressed decays of $D^0$
into two pseudoscalars chosen from $\pi^0$, $\eta$, $\eta'$ that the
corresponding amplitudes depend only on $C'$, $E'$ and $SE'$.  There are five
such decays and one may write down sum rules relating the corresponding
amplitudes. Two such sum rules are as follows:
\bea
4 \sx \, \ca(D^0 \to \pi^0 \eta') - 5 \, \ca(D^0 \to \eta \eta) + 4 \, \ca(D^0
\to \eta \eta') & = & 0~, \\
8 \, \ca(D^0 \to \pi^0 \pi^0) + 4 \st \, \ca(D^0 \to \pi^0 \eta) + 3 \, \ca(D^0
\to \eta \eta) & = & 0~.
\eea
For each sum rule, one can draw a triangle whose sides are given by the
magnitudes of the amplitudes involved in the corresponding sum rule.  Using the
measured values of amplitudes one finds that the angles of such triangles are
non-trivial (i.e., none of them are very near zero or $180^\circ$.)  One may
thus infer that the relevant amplitudes have non-trivial relative strong
phases.

One can also write a sum rule that relates the squares of magnitudes of the
amplitudes instead of the amplitudes themselves:
\beq
8|\ca(D^0 \to \pi^0 \eta')|^2 + 16|\ca(D^0 \to \pi^0 \pi^0)|^2 =
16|\ca(D^0 \to \pi^0 \eta)|^2 +  9|\ca(D^0 \to \eta \eta)|^2~.
\eeq
The magnitudes of the decay amplitudes are well quantified.  The above
relationship thus may easily be tested using the amplitudes from Table
\ref{tab:scskpi} ($D^0 \to \pi^0 \pi^0$) and Table \ref{tab:pluseta} ($D^0 \to
\pi^0 \eta, \pi^0 \eta', \eta \eta$.) In the present case we find
\beq
8|\ca(D^0 \to \pi^0 \eta')|^2 + 16|\ca(D^0 \to \pi^0 \pi^0)|^2 = 325 \pm 33~,
\eeq
\beq
16|\ca(D^0 \to \pi^0 \eta)|^2 +  9|\ca(D^0 \to \eta \eta)|^2   = 440 \pm 39~,
\eeq
in units of $10^{-14}$ GeV$^2$.
Evidently there is little more than a two--sigma deviation from the identity.
This is another signature of deviation from flavor-SU(3) symmetry since one has
already assumed such a symmetry in writing representations for the relevant
decays.

\section{Doubly-Cabibbo-suppressed decays}

In Table \ref{tab:dcs} we expand amplitudes for doubly-Cabibbo-suppressed
decays in terms of the reduced amplitudes $\ttl \equiv - \tan^2 \theta_C T$,
$\tc \equiv - \tan^2 \theta_C C$, $\te \equiv - \tan^2 \theta_C E$, and
$\ta \equiv - \tan^2 \theta_C A$.

% This is Table V
\begin{table}
\caption{Branching ratios, amplitudes, and representations in terms of
reduced amplitudes for doubly-Cabibbo-suppressed decays.
\label{tab:dcs}}
\begin{center}
\begin{tabular}{|c|c|c|c|c|c|} \hline
Meson &    Decay    &     $\b$      & $p^*$ &    $|\ca|$    & Rep. \\
      &    mode     &  $(10^{-4})$  & (MeV) &($10^{-7}$ GeV)&      \\ \hline
$D^0$ &$K^+ \pi^-$&$1.45\pm0.04^{~a}$&861.1 & 1.54$\pm$0.02 & $\ttl+\te$ \\
      & $K^0 \pi^0$ &     $^b$      & 860.4 &     $^b$      & $(\tc-\te)/\s$ \\
      & $K^0 \eta$  &     $^b$      & 771.9 &     $^b$      & $\tc/\st$ \\
      & $K^0 \eta'$ &     $^b$      & 564.9 &     $^b$      & $-(\tc+3\te)/\sx$
\\ \hline
$D^+$ & $K^0 \pi^+$ &     $^b$      & 862.6 &     $^b$      & $\tc+\ta$ \\
      &$K^+ \pi^0$&$2.37\pm0.32^{~a}$&864.0 & 1.23$\pm$0.08 & $(\ttl-\ta)/\s$\\
      & $K^+ \eta$  &     $^c$      & 775.8 &      --       & $-\ttl/\st$ \\
      & $K^+ \eta'$ &     $^c$      & 570.8 &      --       & $(\ttl+3\ta)/\sx$
\\ \hline
$D_s^+$& $K^0 K^+$  &     $^b$      & 850.3 &     $^b$      & $\ttl+\tc$ \\
\hline
\end{tabular}
\end{center}
\leftline{$^a$ Ref.\ \cite{Yao:2006px}.  $^b$ Amplitude involves interference
between DCS process shown and the}
\leftline{corresponding CF decay to $\ok + X$.  $^c$ Studied in Ref.\
\cite{Nisar:2007}.}
\end{table}

With $\tan \theta_C = 0.2317$ one predicts $|\ca(D^0 \to K^+ \pi^-)| = 1.35
\times 10^{-7}$ GeV and $|\ca[D^+ \to K^+(\pi^0,\eta,\eta')] = (0.98,0.86,0.83)
\times 10^{-7}$ GeV.  The experimental amplitudes for $D^0 \to K^+ \pi^-$ and
$D^+ \to K^+ \pi^0$ are, respectively, 14\% and $(26 \pm 8)\%$ above the
flavor-SU(3) predictions.  Ref.\ \cite{Nisar:2007} has demonstrated the
feasibility of testing the predictions for $D^+ \to K^+ (\eta,\eta')$ with
the full CLEO-c data sample.

\subsection{$D^0 \to (K^0 \pi^0, \ok \pi^0)$ interference}

\begin{figure*}
\mbox{\includegraphics[width=0.46\textwidth]{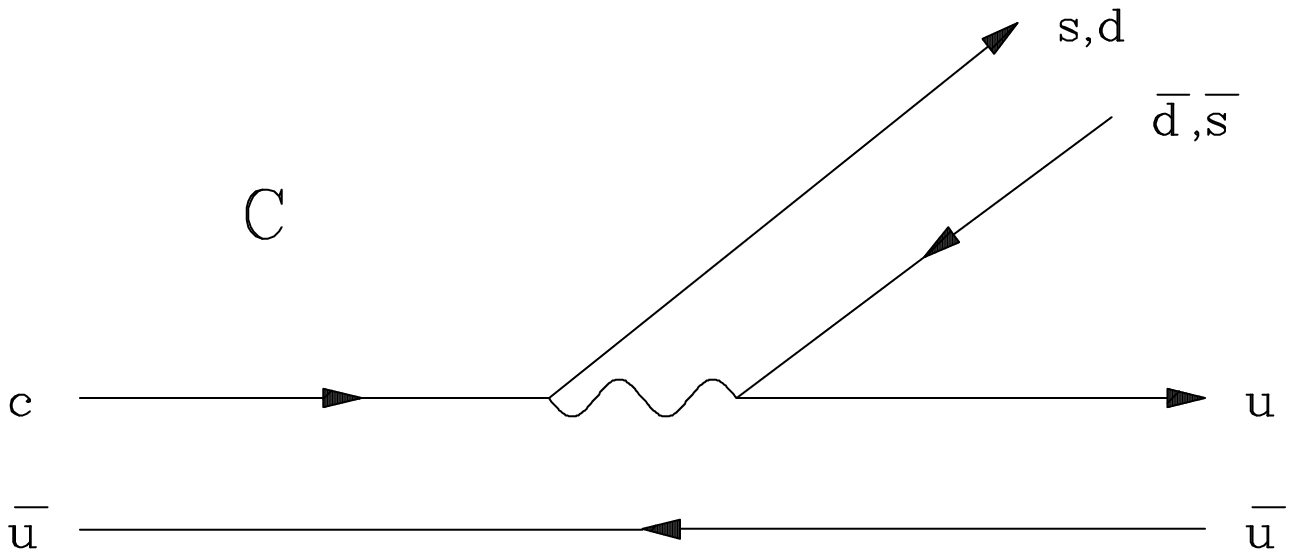} \hskip 0.3in
      \includegraphics[width=0.46\textwidth]{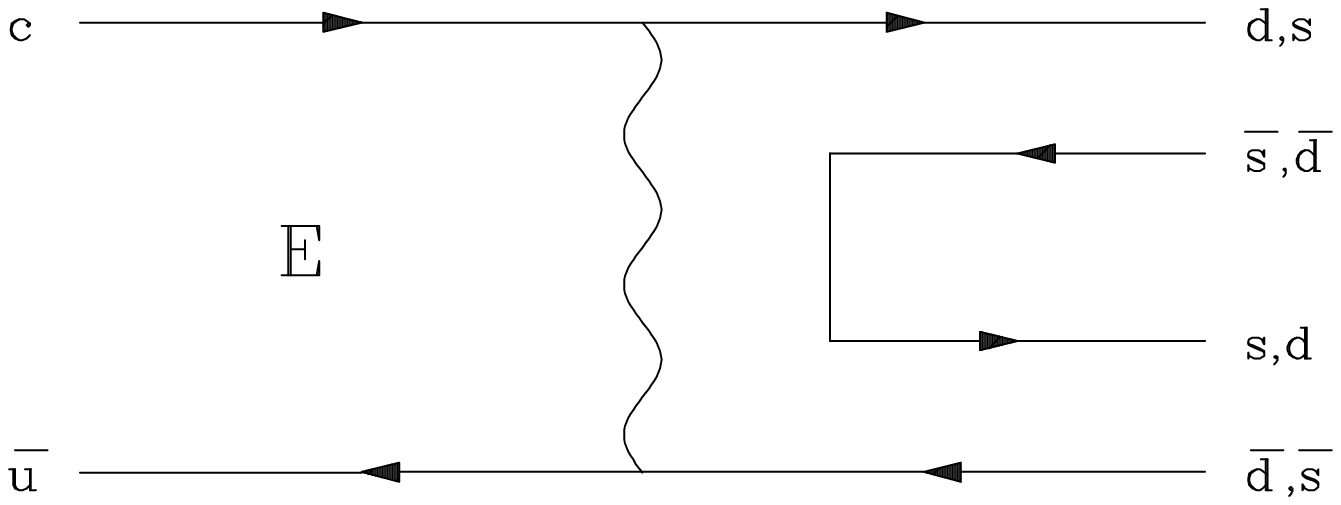}}
\caption{Graphs contributing to $D^0 \to (K^0 \pi^0, \ok \pi^0)$.
\label{fig:Dzint}}
\end{figure*}

The decays $D^0 \to K^0 \pi^0$ and $D^0 \to \ok \pi^0$ are related to
one another by the U-spin interchange $s \leftrightarrow d$, and
SU(3) symmetry breaking is expected to be extremely small in this relation
\cite{Rosner:2006}.  Graphs contributing to these processes are shown in
Fig.\ \ref{fig:Dzint}.

% This is Figure 4

The CLEO Collaboration \cite{He:2007aj} has reported the asymmetry
\beq
R(D^0) \equiv  \frac{\Gamma(D^0 \to K_S \pi^0) - \Gamma(D^0 \to K_L \pi^0)}
 {\Gamma(D^0 \to K_S \pi^0) + \Gamma(D^0 \to K_L \pi^0)}
\eeq
to have the value $R(D^0) = 0.108 \pm 0.025 \pm 0.024$, consistent with
the expected value \cite{Rosner:2006,Bigi} $R(D^0) = 2 \tan^2 \theta_C \simeq
0.107$.  One expects the same $R(D^0)$ if $\pi^0$ is replaced by $\eta$
or $\eta'$ \cite{Rosner:2006}.

\subsection{$D^+ \to (K^0 \pi^+,\ok \pi^+)$ interference}

% This is Figure 5
\begin{figure*}
\mbox{\includegraphics[width=0.31\textwidth]{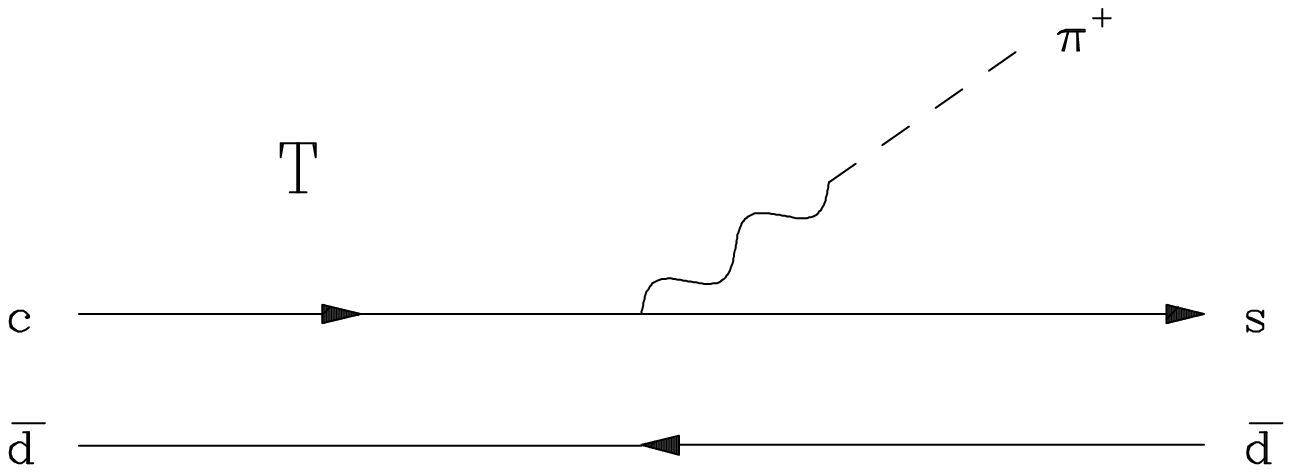} \hskip 0.2in
      \includegraphics[width=0.31\textwidth]{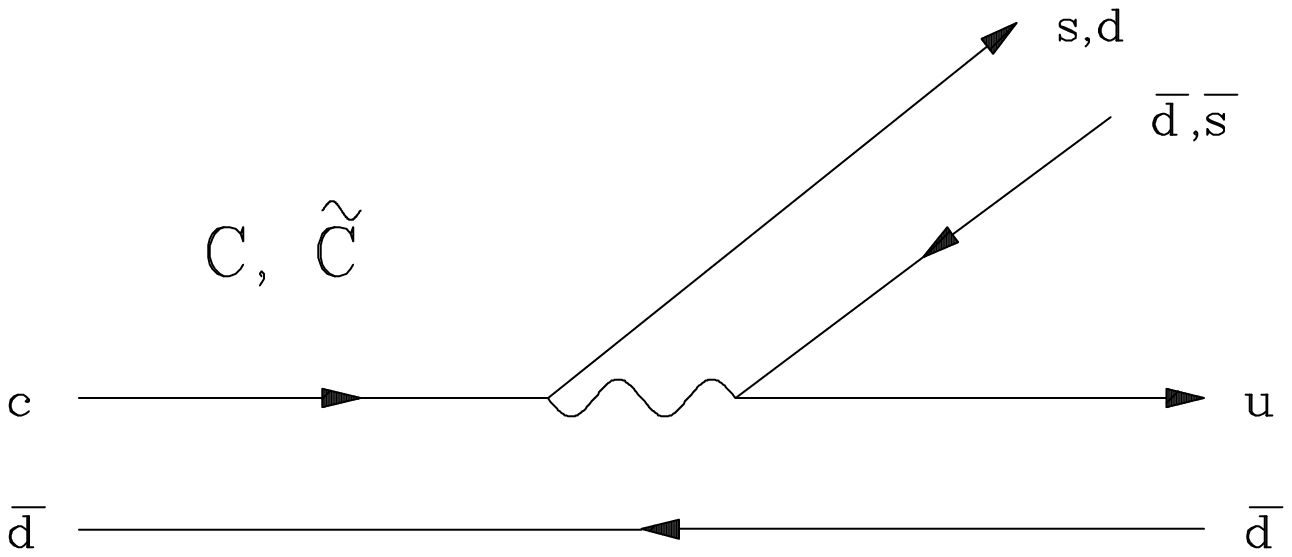} \hskip 0.2in
      \includegraphics[width=0.31\textwidth]{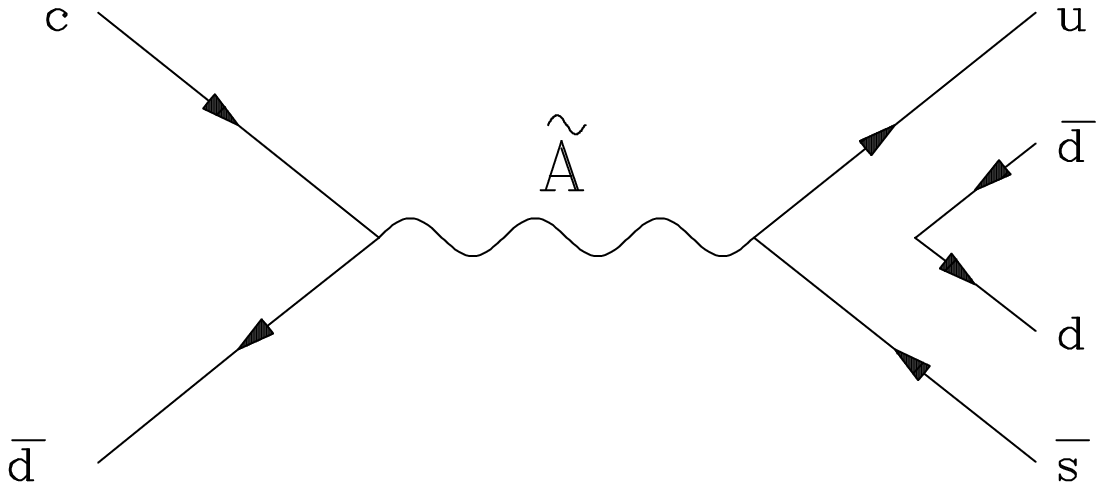}}
\caption{Amplitudes $T$ and $C$ contributing to $D_s^+ \to \ok \pi^+$;
amplitudes $\tc$ and $\ta$ contributing to $D^+ \to K^0 \pi^+$.
\label{fig:Dpint}}
\end{figure*}

In contrast to the case of $D^0 \to (K^0 \pi^0, \ok \pi^0)$, the decays
$D^+ \to (K^0 \pi^+,\ok \pi^+)$ are not related to one another by
a simple U-spin transformation.  Amplitudes contributing to these
processes are shown in Fig.\ \ref{fig:Dpint}.  Although both processes
receive color-suppressed ($C$ or $\tc$) contributions, the Cabibbo-favored
process receives a color-favored  tree ($T$) contribution, while the
doubly-Cabibbo-suppressed process receives an annihilation ($\ta$)
contribution.  In order to calculate the asymmetry between $K_S$ and
$K_L$ production in these decays due to interference between CF and
DCS amplitudes, one can use the determination of the CF amplitudes
discussed previously and the relation between them and DCS amplitudes.
Thus, we define
\beq
R(D^+) \equiv \frac{\Gamma(D^+ \to K_S \pi^+) - \Gamma(D^+ \to K_L \pi^+)}
 {\Gamma(D^+ \to K_S \pi^+) + \Gamma(D^+ \to K_L \pi^+)}
\eeq
and predict
\bea
R(D^+) & = & - 2~{\rm Re}~\frac{\tilde{C} + \tilde{A}}{T+C} \nonumber \\
       & = & 2 \tan^2 \theta_C~{\rm Re}~\frac{C+A}{T+C} \nonumber \\
       & = & -0.006^{~+~0.033}_{~-~0.028} ~,
\eea
where the error is assumed to be dominated by its dominant source, the
uncertainty in $|A|$ (see Fig.\ \ref{fig:cf}).  This is consistent with the
observed value $R(D^+) = 0.022 \pm 0.016 \pm 0.018$ \cite{He:2007aj}.  The
relative phase of $C+A$ and $T+C$ is nearly $90^\circ$, as can be seen
from Fig.\ \ref{fig:cf}.  The real part of their ratio hence is small.
If one uses instead amplitudes based on fitting all CF decays except
$D_s^+ \to \ok K^+$, as in Ref.\ \cite{Chiang:2003}, one predicts instead
$R(D^+) = 0.013 \pm 0.035$.

A similar exercise can be applied to the decays $D_s^+
\to K^+ K^0$ and $D_s^+ \to K^+ \ok$, which are related by U-spin to the
$D^+$ decays discussed here.  The corresponding ratio
\beq
R(D_s^+) \equiv \frac{\Gamma(D_s^+ \to K_S K^+) - \Gamma(D_s^+ \to K_L K^+)}
 {\Gamma(D_s^+ \to K_S K^+) + \Gamma(D_s^+ \to K_L K^+)}
\eeq
is predicted to be
\bea
R(D_s^+) & = & - 2~{\rm Re}~\frac{\tilde{C} + \tilde{T}}{A+C} \nonumber \\
       & = & 2 \tan^2 \theta_C~{\rm Re}~\frac{C+T}{A+C} \nonumber \\
       & = & -0.003^{~+~0.019}_{~-~0.017}~.
\eea
Using amplitudes based on all CF decay rates except that for $D_s^+ \to
\ok K^+$, one predicts instead $R(D_s^+) = 0.005 \pm 0.017$.

\section{Other theoretical approaches}

One can invoke effects of final state interactions to explain arbitrarily
large SU(3) violations (if, for example, a resonance with SU(3)-violating
couplings dominates a decay such as $D^0 \to \pi^+ \pi^-$ or $D^0 \to K^+
K^-$).  As one example of this approach \cite{Buccella:1996}, both resonant and
nonresonant scattering can account for the observed ratio $\Gamma(D^0 \to
K^+ K^-)/\Gamma(D^0 \to \pi^+ \pi^-) = 2.87 \pm 0.08$.  This same approach
predicted $\b(D^0 \to K^0 \ok) = 9.8 \times 10^{-4}$, a level of SU(3)
violation consistent with the world average of Ref.\ \cite{Yao:2006px} but
far in excess of the recent CLEO value \cite{Bonvicini:2007}.  The paper of
Ref.\ \cite{Buccella:1996} may be consulted for many predictions for $PV$ and
$PS$ final states in charm decays, where $V$ denotes a vector meson and
$S$ denotes a scalar meson.  Results for $PV$ decays also may be found in
Refs.\ \cite{Rosner:1999,Chiang:2003,Chiang:2002,Cheng:2003}.

The recent discussion of Ref.\ \cite{Gao:2007} entails a prediction
$A \simeq - 0.4E$, essentially as
a consequence of a Fierz identity and QCD corrections.  Tree amplitudes are
obtained from factorization and semileptonic $D \to \pi$ and $D \to K$ form
factors.  The main source of SU(3) breaking in $\ttl/T$ is assumed to come
from $f_K/f_\pi = 1.22$.  Predictions include asymmetries $R(D^{0,+}) =
(2 \tan^2 \theta_C,~0.068 \pm 0.007)$, and -- via a sum rule for $D^0
\to K^\mp \pi^\pm$ and $D^+ \to K^+ \pi^0$ -- a prediction of the relative
strong phase $\delta$ between $D^0 \to K^+ \pi^-$ and $D^0 \to K^- \pi^+$,
$|\delta| \simeq 7$--$20^\circ$ (to be compared with 0 in exact SU(3) symmetry
\cite{SU3}).

\section{Summary}

We have shown that the relative magnitudes and phases of amplitudes
contributing to charm decays into two pseudoscalar mesons are describable by
flavor symmetry.  We have verified that there are large relative phases
between the color-favored tree amplitude $T$ and the color-suppressed
amplitude $C$, as well as between $T$ and $E$.  The phase of $A$ is nearly
opposite to that of $E$, as originally found in Ref.\ \cite{Rosner:1999},
but its magnitude is only about 1/3 that of $E$, whereas it was nearly
that of $E$ in Refs.\ \cite{Rosner:1999} and \cite{Bhattacharya:2007jc}.  The
difference is due primarily to new measurements of absolute branching ratios
for Cabibbo-favored (CF) $D_s$ decays by the CLEO Collaboration
\cite{Alexander:2008}.

The largest symmetry-breaking effects are visible in singly-Cabibbo-suppressed
(SCS) decays, particularly in the $D^0 \to(\pi^+ \pi^-/K^+ K^-)$ ratio  which
are at least in part understandable through form factor and decay constant
effects.  Decays involving $\eta$, $\eta'$ are mostly describable with
small ``disconnected'' amplitudes, a possible exception being in SCS $D^+$ and
$D_s^+$ decays.

One sees evidence for the expected interference between Cabibbo-favored and
doubly-Cabibbo-suppressed (DCS) decays in $D^{0,+} \to K_{S,L} \pi^{0,+}$
decays.  This interference leads to a measurable rate asymmetry in the decays
$D^0 \to K_{S,L} \pi^0$ but none in $D^+ \to K_{S,L} \pi^+$.

\bigskip
\section*{ACKNOWLEDGMENTS}
J.L.R. wishes to thank C.-W. Chiang, M. Gronau, and Z. Luo for enjoyable
collaborations on some earlier aspects of the results described here.  We are
grateful to S. Blusk, H. Mahlke, A. Ryd, and E. Thorndike for helpful
discussions.  This work was supported in part by the United States Department
of Energy through Grant No.\ DE FG02 90ER40560.

\end{document}